\documentstyle[a4wide,amssymb,graphicx,aps]{revtex}

\begin{document}

\title{Comment on ``Formation of a Dodecagonal Quasicrystalline Phase
  in a Simple Monatomic Liquid''}

\author{J. Roth, \\
Institut f\"ur Theoretische und Angewandte Physik\\
Universit\"at Stuttgart\\
Pfaffenwaldring 57\\
70550 Stuttgart, Germany}

\maketitle

PACS No 61.44.Br, 61.20.Ja, 64.60.Qb

In a recent paper M.~Dzugutov\cite{dzugutov92} describes a molecular
dynamics cooling simulation where he obtained a large monatomic
dodecagonal quasicrystal from a melt. The structure
was stabilized by a special potential\cite{dzugutov90} designed to
prevent the nucleation of simple dense crystal structures. In this
comment we will give evidence that the ground state structure for
Dzugutov's potential is an ordinary bcc crystal. More detailed
publications of the phase diagram of the Dzugutov potential are in
preparation\cite{denton97,roth97}.

The results are obtained by molecular dynamics simulation. 
The constraint method\cite{alltil} has been applied to simulate NVT or NPT
ensembles with an extension that permits constant
temperature or pressure gradients also. The equations of motion are
integrated by a fourth-order Gear algorithm.
The time increment was $\delta t$=0.005 (all quantities are given in
Lennard-Jones units). 

The initial structures have been obtained by melting and equilibrating
crystals with 250, 500 or 1024 atoms at T=4.0 and densities of
$\rho=0.63$, 0.55 and 0.857 , resp.\ for 10$^5$ time steps at constant volume.
The 250 atom sample has been cooled at rates of -0.002/$\delta t$ and
-0.001/$\delta t$. With the faster rate one obtains a tetrahedrally
close packed (tcp) structure\cite{shoemaker90} where all atoms are 12,
14 or 15 fold coordinated and surrounded by Frank-Kasper
polyhedra. The quasicrystal also belongs to this class. The bcc
structure is tetrahedrally packed to, but the coordination
polyhedron is not of the Frank-Kasper type. The radial density 
functions of all the structures are quite similar, although the first
maximum differs in shape between bcc and tcp in a good quality
sample.
Cooling the 500 atom system at -0.001/$\delta t$ and the
1024-atom system at -0.00025/$\delta t$ also leads to bcc crystals. At
faster rates the large system shows partial ordering, but the
structure as a whole did not form an ordered crystal. 
Since 250 and 1024 are magic numbers for a bcc
crystal one could assume that these samples get locked into the
bcc structure. This is certainly not the case due to several reasons:
first, the nucleating structures contain vacancies, which change the
required number of atoms to a non-magic one. Second, the samples are
sometimes twinned or contain more than one domain. Third, the basic [100]
and equivalent lattice planes are {\em never} parallel to the
simulation box, and the nucleated solids often contain screw
dislocation loops and are globally twisted, but locally the atoms
are in a bcc environment.

Cooling the fluid at high {\em constant pressure} always yields a bcc
structure which sometimes evolves into a fcc structure at low enough
temperature. At P=5 a tcp structure is obtained at -0.002/$\delta t$
and a bcc crystal at -0.001/$\delta t$. If the fluid is compressed at
constant temperature the results are similar. In the range from T=2.5
to 1.0 the bcc structure forms at a compressing rate of 0.1/$\delta
t$. At T=0.6 a twinned crystal was found with a few tcp atoms. 

It has been shown by density functional calculations\cite{denton97}
and energy minimization calculations\cite{roth97} that the bcc
structure is the lowest in energy at densities 
where the pressure is low. At increasing density the tcp
$\sigma$-phase\cite{shoemaker90} (a low order approximant of
the quasicrystal) becomes more stable, and at high
density the fcc structure is the most stable. The difference between
bcc and $\sigma$-phase is rather small so that it is currently not
clear which one is more stable at low and moderate pressure and $T > 0$.

\begin{figure}
\caption{Bond angle correlation function for nearest neighbour atoms
The histogram gives the freqeuncy of angles as a function of the
cosine. The liquid (full line) is similar to a tcp structure (dotted
line), whereas the perfect bcc structure is completely different, but
close to the nucleated structure (dash-dotted line).
\label{bac}}
\vspace*{1cm}
\begin{center}
\includegraphics[width=12cm,angle=0]{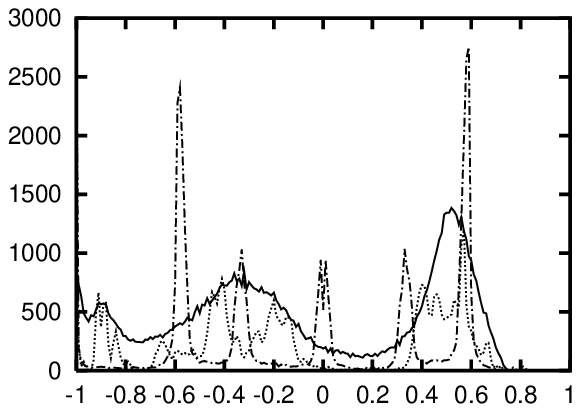}
\end{center}
\end{figure}

At a high cooling rate one obtains an
glass. In an intermediate range at low pressure a
tetrahedrally close packed structure or a quasicrystal is
favoured since its structure is close to the fluid structure (Fig.\
\ref{bac}). At slow cooling rates the system has 
enough time to nucleate into the stable bcc structure. With increasing
system size it takes more and more time for the fluid to reorder into
the bcc structure, therefore the tcp structure will be generated.

\end{document}